# Improving the Performance of WLANs by Reducing Unnecessary Active Scans

Dheryta Jaisinghani, *Member, IEEE,* Vinayak Naik, *Member, IEEE,* Sanjit Kaul, *Member, IEEE,*
Rajesh Balan, *Member, IEEE,* and Sumit Roy, *Fellow, IEEE*

*Abstract*—We consider the problem of excessive and unnecessary active scans in heavily utilized WLANs during which low rate probe requests and responses are broadcast. These management frames severely impact the goodput. Our analysis of two production WLANs reveals that lesser number of non-overlapping channels in $2.4$ GHz makes it more prone to the effects of increased probe frames than $5$ GHz. We find that not only up to **90%** of probe responses carry redundant information but the probe traffic can be as high as $60\%$ of the management traffic. Furthermore, active scanning severely impacts real-time applications at a client as it increases the latency by $91$ times.
We present a detailed analysis of the impact of active scans on an individual client and the whole network. We discuss three ways to control the probe traffic in production WLANs – access point configurations, network planning, and client modification. Our proposals for access point configuration are in line with current WLAN deployments, better network planning is device agnostic in nature, and client modification reduces the average number of probe requests per client by up to $50\%$ without hampering the ongoing WiFi connection.

## I. INTRODUCTION

When designing a WiFi network, network designers typically focus on placing access points (APs) at the locations needed to satisfy the required average throughput per user. They then iterate through this process using numerous traffic patterns and usage models until they reach a final network design that can achieve the required throughput levels with the minimum number of APs (to reduce costs). However, in practice, there are additional factors that are not currently considered in this design process that can and do have a severe impact on the available bandwidth of a WiFi network.

In this paper, we present a detailed study, using two different production WiFi networks, that show the impact of one of these previously non-considered factors – namely, the impact of excessive broadcast probe requests (PReqs) and probe responses (PRes's) on the available network throughput in the production WiFi networks.

As part of the `802.11` standard, WiFi clients send broadcast PReqs, across a wide range of channel frequencies, to discover APs in the environment [1]. Usually, every AP hearing these PReqs responds with a unicast PRes that contains information about that network. This process is commonly referred to as Active Scanning. Note: There are many exceptions to when and how APs respond to PReqs, which we will list in Section II. Normally, these PReqs and PRes's do not cause any significant throughput issues as they comprise a very small percentage, $< 2\%$ across all channels, of the overall network management traffic. However, we have found that in the heavily utilized networks, where channels are at least $50\%$ utilized, the amount of probe traffic grows up to $50\%$, at least, of the total management traffic in just a second. This, in turn, brings down the goodput exponentially [2]. The reason for this dramatic drop in goodput is because probe traffic, as per the WiFi specifications to ensure the highest delivery probability is sent at full power, at the *lowest bit rate*, and on most channels. This low bit rate increases their transmission time and during this time, other clients are unable to send or receive data on the shared wireless spectrum. Thus, if these slow frames start increasing beyond a certain point, the overall efficiency of the network starts degrading as the clients and APs are unable to find free spectrum slots to send and receive required data.

To understand this phenomenon better, we conducted a detailed study of a campus network in India that primarily operates on the $2.4$ GHz spectrum and then conducted a follow on study of a campus network in Singapore that primarily operates on the $5$ GHz spectrum. We found that the performance of $2.4$ GHz networks is greatly affected by excessive probe traffic while the impact was negligible on $5$ GHz networks. Number of PReqs in $5$ GHz were found to be $1.65$ times lesser than $2.4$ GHz. The fundamental reason for this was that congestion in the $2.4$ GHz network was higher as there are only $3$ non-overlapping channels available. Thus a large number of clients were consistently sending PReqs as they moved around the campus and when they felt that the available throughput was not high enough [3].

Even though the $5$ GHz network was mostly unaffected by this phenomena, as there are multiple non-overlapping channels in this frequency band to isolate every AP from its neighbors, we believe that the results in this paper are still very useful to network designers as $2.4$ GHz networks are still very common in Asia, Africa, and other parts of the world due to the lower cost (for both APs and clients) and higher range of these networks. Unofficially, it is estimated that $\approx 85\%$ of the devices in India run on the $2.4$ GHz spectrum. In addition, it is likely that emerging IoT networks that decide to use WiFi as the wireless medium will use the $2.4$ GHz band primarily

---

D. Jaisinghani, V. Naik, and S. Kaul are with IIIT-Delhi, New Delhi India
E-mail: {dherytaj,naik,skkaul}@iiitd.ac.in
R. Balan is with Singapore Management University, Singapore
E-mail: rajesh@smu.edu.sg
S. Roy is with University of Washington-Seattle, WA, USA
E-mail: roy@ee.washington.edu

for price and range reasons.

**Contributions**

1) We compare the active scanning behavior of clients in 2.4 GHz and 5 GHz in a live network. We find that 5 GHz WLANs with near-stationary WiFi clients see lesser PReqs than 2.4 GHz.[Section III]
2) We provide a detailed analysis on *why* and *how* does active scanning affect the performance of a WLAN. We pursue these questions separately from the perspective of an individual client as well as whole network. For a client, we report the impact of active scanning on the latency and if it enables the client to choose a different AP.
   For the network-wide scale, we analyze three real-world WiFi networks to understand the extent of probe traffic. We report numbers for the amount, frame sizes, data rates, inter-frame arrival times, and airtime utilization of the probe traffic. We introduce a new metric – *Redundant Probe Traffic*, to measure the amount of redundant information fetched by PRes's. Finally, we present a case study to demonstrate the effect of probe traffic on goodput in a heavily utilized network.[Section IV]
3) We present three ways to control the growth of probe traffic [Section V]. We study many existing WiFi driver implementations and user space applications to arrive at three causes of active scanning – *Discovery*, *Connection Establishment*, and *Connection Maintenance*. We develop a sniffer-based, device agnostic, inference mechanism for detecting these causes in real WiFi networks. With the proposed inference mechanism, we identify the extent of active scanning, and the related causes in our datasets.[Section VI].
   We develop a modified scanning strategy that dynamically chooses when to trigger active scanning. We implement this strategy in clients with different WiFi chipsets and evaluate their performance in real and uncontrolled WLAN for 9 hours. The modified scanning strategy reduces the average number of PReqs per WiFi client by up to 50% without impacting the ongoing connection.[Section VII].

We discuss the related works in Section VIII and conclude in Section IX.

## II. PROCEDURE OF ACTIVE SCANNING

In this section, we explain the procedure of active scanning from the holistic view of a WiFi client.

Active scanning is either triggered in the userspace or the kernel space of an OS; Figure 1, shows the interaction among different layers of both spaces and with the WiFi chipset of a client. Examples of user space applications include the Ubuntu Network Manager and the Android `wpa_supplicant`. The kernel space includes the WiFi MAC drivers such as `mac80211` or device drivers such as `ath9k` and `iwlwifi`. Irrespective of the location from where active scanning is triggered, when it starts, client prepares a PReq frame. If SSID field in it is empty, the request is known as a Null Probe Request otherwise it is known as an Individually Addressed or Directed Probe Request. Such a frame is sent to search an AP to which client must have associated in the past. The client scans a list of channels for each frequency band as per its configuration. For each channel in the list, the client sends a PReq. Type and number of PReqs depend on the code which triggered active scanning. For example, `wpa_supplicant` in Android sends directed PReqs to the past SSIDs.

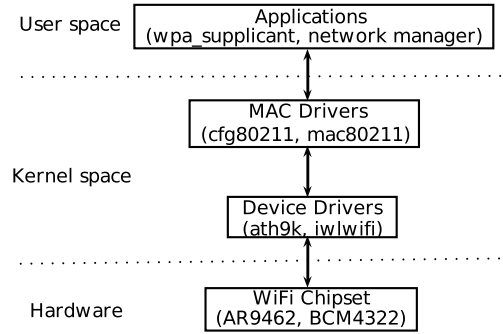

Fig. 1: The different layers of WiFi functionality.

The APs which could listen to PReqs, respond with PRes's addressed to the client. On receiving all PRes's, it sends ACK to all the APs, which responded. In case the ACK is not received by the AP, it resends the PRes. Ideally, all APs in the vicinity of a client can respond, however in the production networks, for example, Cisco/Aruba WLANs, APs may decide to or not to respond. Their decision depends on factors such as load balancing algorithms, band steering, and various thresholds as set by the WLAN administrator such as the number of associated clients, number of PReqs to respond to, and RSSI for a client.

Though, active scanning seems to be a simple process of sending and receiving multiple PReqs and PRes's, respectively; however, in practice it is not only a result of various causes defined at layers of software stack but also complex interactions among these layers. Diversified user space applications and kernel space drivers with the availability of many OS's and WiFi chip vendors makes active scanning even more complicated. Therefore, an in-depth analysis is required.

In the next section, we compare the scanning behavior of devices in 2.4 GHz and 5 GHz.

## III. COMPARING ACTIVE SCANNING– 2.4 GHz VS 5 GHz

Most WiFi devices today support dual band operation, *i.e.*, 2.4 GHz and 5 GHz. Both these bands exhibit different channel characteristics and interference patterns. Thus, we compare the active scanning behavior of clients in both these bands.

### A. Data Overview

We collected WiFi traffic at one floor in one of the buildings of Singapore Management University (SMU) for 6 hours, from 11 AM to 5 PM. This time slot coincides with the office hours, thereby ensuring the recording of traffic from maximum clients. This floor has 15 APs deployed with each AP broadcasting 4 SSIDs. At least 200 people, including students, faculty, staff, visitors, with a minimum of 1 WiFi device/person are expected throughout the day. We recorded frames with sniffers on channels 1, 6, 11 in 2.4 GHz and 36, 52, 161 in 5 GHz. Our sniffers were TP-Link WN721N USB WiFi adapters for 2.4 GHz and Intel Centrino 6230 PCI cards for 5 GHz. We only examined the PHY header and type of frames, specifically PReqs for this analysis. Sniffer recorded a total of 222 and 49 clients/minute in 2.4 GHz and 5 GHz, respectively. Out of these 70 clients and 20 clients in 2.4 GHz and 5 GHz, respectively, are found to be associated. We assume the remaining clients to be unassociated. However, please note that the number of unassociated clients is an overestimate because of MAC address randomization [4].

*B. Data Analysis*

**Hypothesis** We hypothesize that the stationary clients in 5 GHz transmit lesser PReqs than the mobile clients. The reason is reduced interference in 5 GHz, which in turn reduces the amount of PReqs due to the *Connection Maintenance* cause. In the other case, the lower range of 5 GHz frequency band results in a dense deployment of APs to ensure coverage. Hence, the mobile clients trigger active scanning as soon as the RSSI falls below roaming threshold, which is $-70$ dBm typically. Thus, the mobile clients experience frequent handovers and probe more in 5 GHz.

We compare the scanning behavior in both frequency bands for the following metrics – ($a$) #PReqs transmitted by each WiFi client/minute and ($b$) Effect of RSSI variations.

Our first metric – #PReqs/client/minute signifies the normalized value of the number of PReqs transmitted by associated and unassociated clients together in each frequency band. Figure 2 presents an analysis of this metric. We notice that while #PReqs/client/minute in 2.4 GHz are limited to 1 or 2, in 5 GHz as high as 6 PReqs/client/minute are transmitted. Not only this but the variance is also 5 times higher than 2.4 GHz. With this, it appears that 5 GHz will experience higher probe traffic than 2.4 GHz. However, as we dwell deeper in the analysis, we find that this is a fallacy.

We proceed further to analyze the effect of RSSI on the #PReqs, #Status Changes, and the type of SSIDs. All results are normalized to 500 clients/minute in each band. We infer the RSSI values from PHY header of data frames recorded at the sniffers. Figure 3 shows the result of this analysis. We considered 8 classes of RSSI from $-20$ dBm to $-90$ dBm with decrementing $-10$ dBm/class. For every class, we plot the average #PReqs transmitted. While 2.4 GHz sees almost constant #PReqs for all RSSI classes, 5 GHz shows sudden growth in #PReqs as RSSI is close to roaming threshold (from $-40$ to $-70$ dBm). In such scenario, #PReqs in 2.4 GHz grow by 3.15% while in 5 GHz they grow by 209.02%. For all other classes, either there are no PReqs in 5 GHz or $<=$ 2.4 GHz. To verify this, we study the effect of variation in RSSI on the number of times clients change their association status. We term this metric as #Status Changes. Figure 4 shows the result of this analysis. While in 2.4 GHz, #Status Changes are mostly constant across RSSI classes, in 5 GHz we notice 64.28% increase as the RSSI decrease from $-40$ to $-70$ dBm.

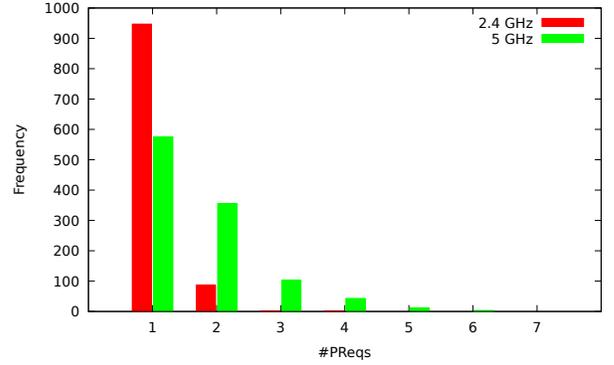

Fig. 2: Frequency of #PReqs/client/minute in 2.4 GHz and 5 GHz. Frequency of 1 PReq/client/minute if 1.65 times lesser for 5 GHz than 2.4 GHz. Average and maximum are 1 and 2 in 2.4 GHz while 2 and 6 in 5 GHz. Standard deviation and Variance are 0.40 and 0.16 in 2.4 GHz while 0.93 and 0.87 in 5 GHz.

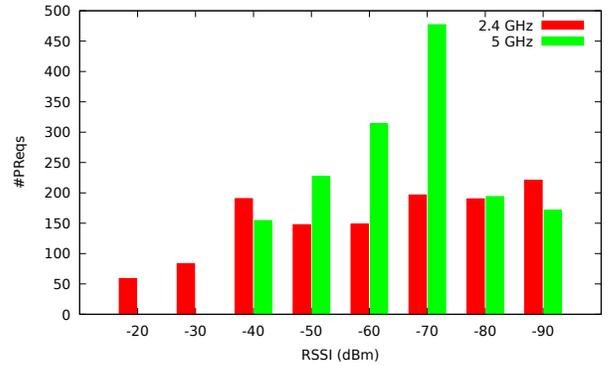

Fig. 3: Comparing the #PReqs/minute with the variation in RSSI. Notice the growth in #PReqs in 5 GHz as RSSI approaches close to $-70$ dBm.

We validate this result by analyzing the type of SSIDs broadcast in the PReqs when RSSI is close to roaming threshold. We categorize the SSIDs as enterprise and non-enterprise SSIDs. Enterprise SSIDs are the ones, which are part of WLAN deployment, while non-enterprise SSIDs are not. The Occurrence of such PReqs signifies triggering of handover at the client. Our analysis reveals that the number of enterprise SSIDs in the PReqs grow from 46.66% at $-40$ dBm to 81.23% at $-70$ dBm in 5 GHz, while in 2.4 GHz they stay at around 45%.

This result provides conclusive evidence for the argument that more #PReqs in 5 GHz are due to mobile WiFi clients. However, stationary clients send lesser PReqs in 5 GHz. Revisiting Figure 2, we now understand that 5 GHz experiences 1.65 times lower instances of 1 PReq/client/minute than 2.4 GHz. Clients responsible for this number were near-stationary and associated. However, high variance in #PReqs/client/minute is due to high RSSI variations.

Data analysis reveals that stationary clients transmit lesser

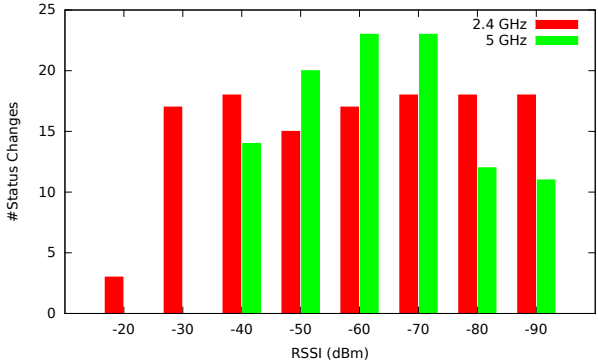

Fig. 4: Comparing the #Status Changes (Associated to Unassociated and vice-versa) with the variation in RSSI. Notice the growth in #Status Changes in 5 GHz as RSSI approaches close to $-70$ dBm.

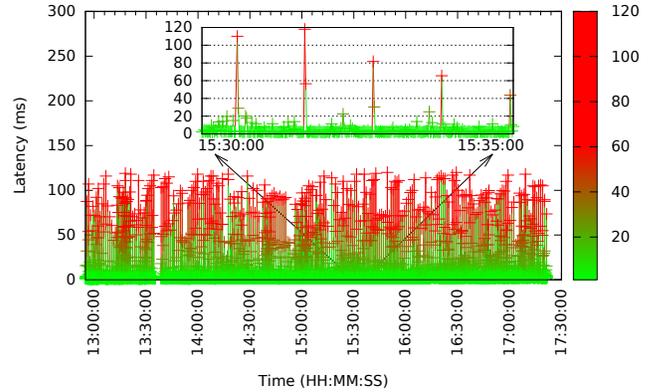

Fig. 5: Impact of active scanning on latency experienced by the client. The plot shows data of $5.5$ hours long experiment. Red peaks denote active scanning. Due to many data points, we zoom in the plot for a 5 minute window. Active scanning is triggered 5 times in this window, notice that on its occurrence latency increases from $\approx 20$ ms to $\approx 100$ ms.

PReqs in 5 GHz band than in 2.4 GHz. Further, as mentioned in Section I most devices even today operate in 2.4 GHz motivates us to study the active scanning in detail. We present the impact of active scanning in the next section.

## IV. IMPACT OF ACTIVE SCANNING

Now, we will understand *why* and *how* does active scanning affect network performance? We pursue this question from two perspectives – (*a*) client-side (Section IV-A) and (*b*) network-wide (Section IV-B). It is important to consider two different perspectives because not only does active scanning introduces latency to an ongoing communication at a client but when it starts, low data rate probe traffic, comprising of PReqs and PRes's, is injected into the network. Low data rate probe traffic consumes airtime that ultimately affects the performance of other clients in the network which are not involved in active scanning [2], [3]. While existing studies aim to reduce scanning delays [5]–[11], the impact of active scanning on communication delays between an AP and a client in dense WLANs remain unknown. Furthermore, the extent of airtime consumed by probe traffic in dense WLANs is unknown. The fact that most of this probe traffic is avoidable makes it even more necessary to understand the impact of active scanning on the performance of a network. We show the drop in goodput due to increase in probe traffic with the help of a case study in sub-section IV-C.

### A. Client-side Perspective

*1) Latency:* To understand the client-side impact, we study how latency, *i.e.*, the time it takes for message transfer between a client and an AP is affected when active scanning is triggered. Latency is an important metric [12] to understand this impact because it ultimately affects real-time applications, such as VoIP.

**Measuring the latency** Existing sniffer-based logs do not allow us to understand latency experienced by a client at the application layer. Therefore, we collect latency data at a stationary client in a dense WLAN deployed in [IIIT-D], where the client receives beacons from nearly 10 BSS's (Basic Service Set). This approach allows us to understand the impact of active scanning on latency in a real network. The client remains associated with one AP throughout the data collection period. Any external applications, e.g. Network Manager, that trigger active scanning are disabled. We instrument the device to trigger active scanning once a minute. Note that this experiment is to study the impact of active scanning on latency experienced by a stationary client, irrespective of its cause. We measure latency with the metric Round Trip Time (RTT) [12] reported by `ping` [13] utility at the client.

The client under observation is a $802.11n$ enabled Atheros AR9462 WiFi chipset and it runs Ubuntu 14.04 OS. We disabled the power-save feature at the client, to rule out the possibility of it affecting the latency. All APs in [IIIT-D] WLAN are $802.11ac$ enabled. Both the client and the APs are capable of operating in both 2.4 GHz and 5 GHz frequency bands. We initiated a `ping` session for 5.5 hours on the client, where it pings the AP to which it is associated. We record RTT values on the WiFi client itself.

**Analysis** Figure 5 shows variation in latency experienced with time. Peaks in latency, by up to 91 times, arise due to active scanning triggered at the client. Active scanning not only impacts latency the moment it is triggered but continues to increase latency for next few seconds after that. We demonstrate this phenomenon in the zoomed-in plot. It happens so because the arrival of data frames at the client result in deferring the transmission of remaining PReqs from the ongoing scan.

*2) Association Patterns:* Now we analyze if active scanning helps in choosing a different BSS. We do this by studying the association pattern of stationary WiFi clients. Precisely, we find a set of BSSs that respond to PReqs broadcast by a WiFi client and another set to which it associates. We perform this analysis in the WiFi network deployed in [IIIT-D]. A typical WiFi client is in the vicinity of 2-3 APs, where each AP broadcasts 5 BSSIDs (Basic Service Set Identifiers). Note that each BSS in operation transmits its beacons and PRes's. We randomly pre-selected 34 clients and analyzed their association patterns for 10 days. While we ensured that the devices should be present on all days of data collection, we did not mandate their association choice.

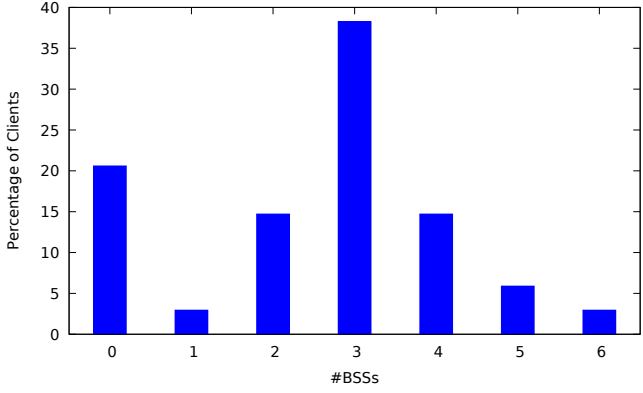

Fig. 6: Association pattern of 34 pre-selected clients in [IIIT-D] over a period of 10 days. 38% of clients selected only 3/11 discovered BSSs.

**Observations** Figure 6 shows the results. All BSS's in operation respond to PReqs without considering the type of client. To exemplify, a rogue client, one who is not authorized to access WLAN, and an official client, one who is entitled to access WLAN, both receive PRes's. Our data analysis confirmed this observation with a maximum of 11 BSS's discovered/WiFi client. Further, APs broadcast many BSSIDs to implement VLANs. Not all authorized clients are allowed to access all VLANs. However, irrespective of this fact, all BSS's respond to all PReqs. We find that more than half of the clients chose maximum 4 out of 11 discovered BSS's to associate. Specifically, 38% associated with 3 while 24% with one of 4 BSS's. Furthermore, 20% did not associate with any BSS. PRes's from BSS's to which clients did not associate were unnecessary.

For stationary clients, active scanning increases the latency. Moreover, even though they discover many BSS's, they rather associate with a much smaller subset. This observation implies that most of the PReqs and PRes's are inessential and their discontinuation should improve performance of network.

Next, we analyze the impact of active scanning from the perspective of the whole network.

### B. Network-wide Perspective

To study the impact of active scanning on a network-wide scale, we consider sniffer-based logs from three real-world WiFi networks. These are [SIGCOMM] [14], [IIT-B] [15], and [IIIT-D]. Table I summarizes details of these datasets. We begin with providing the details of probe traffic in these logs followed by measuring the amount of redundant probe traffic.

*1) Extent of probe traffic in Real-world WiFi Deployments:*
We extract the details about PReqs and PRes's from the logs.

TABLE I: Details of datasets

| # | [SIGCOMM] | [IIT-B] | [IIIT-D] |
|---|---|---|---|
| Band | 5 GHz | 2.4 GHz | 2.4 GHz |
| Days | 2 | 1 | 9 |
| Hours | 14 | 2.5 | 63 |
| Total Frames | 5,922,772 | 40,080,792 | 149,879,321 |
| Management Frames/minute | 468 | 1469 | 5541 |
| Probe Frames/minute | 70 | 761 | 973 |

Particularly, we study the amount of management and probe traffic, frame sizes, physical layer (PHY) data rates, inter-frame arrival times, and airtime utilization [16].

**Amount of management and probe traffic** Table I lists the #frames recorded/minute. We report the numbers for total traffic in general, management and probe traffic in particular. [SIGCOMM] has the lowest #probe frames/minute followed by [IIT-B] and [IIIT-D]. We explain this observation with the following reasons – ($i$) Clients are less likely to probe in 5 GHz band, as discussed in Section III, which is the case for [SIGCOMM]. ($ii$) [SIGCOMM] has the lower #frames and losses recorded than [IIT-B] [15]. ($iii$) [IIIT-D] has the highest #clients and APs recorded, which means a higher #PReqs and PRes's.

As per the `IEEE` standard, frames like beacons, association requests, and responses, authentication requests and responses, and probe traffic together constitute management traffic. We note that out of the total management traffic recorded, the proportion of probe traffic for [SIGCOMM], [IIT-B], and [IIIT-D] is 14.95%, 51.80%, and 17.56%, respectively. These numbers signify that [SIGCOMM] and [IIIT-D] see more beacons and association related events than [IIT-B], which sees more scanning related events.

**Frame Sizes** We observe that PReqs have variable size, but the size of PRes's is near constant for a given WLAN. The reason of which is that PReqs and PRes's are transmitted by clients and APs, respectively. For a given WLAN, specifically the ones deployed in enterprises, clients are from different device vendors while APs are from the same vendor. Figure 7 shows the percentage of PReqs and PRes's for each unique frame size.

The median size of PReqs for each of the datasets is– [SIGCOMM]: 50 bytes, [IIT-B]: 110 bytes, and [IIIT-D]: 70 bytes. More than 80% of PRes's for [SIGCOMM] and [IIT-B] are of 100 bytes and 150 bytes, respectively. However, [IIIT-D] contains PRes's of variable sizes, ranging from 250-300 bytes. The reason for this variability is twofold. First, [IIIT-D] has the higher #BSS's than the other two datasets. #BSS's in [IIIT-D] are 5, in [SIGCOMM] are 2, and in [IIT-B] is 1. Second, a mix of enterprise and non-enterprise APs operate together here. Revisiting the definition of enterprise and non-enterprise, discussed in Section III, non-enterprise APs as the ones that are not part of official WLAN deployment. In the case of [IIIT-D], enterprise APs are deployed as part of Cisco WLAN, while few users deploy non-enterprise APs for their personal usage.

**Data Rates** We find that $\approx 100\%$ of the PReqs are sent at 1 Mbps if the frequency band of operation is 2.4 GHz, as seen in [IIT-B] and [IIIT-D]. However, the rate is 6 Mbps in 5 GHz, as seen in [SIGCOMM]. Data rate of PRes's is found to vary among 1, 6, or 24 Mbps. For each dataset the data rate of PRes's is as follows - [SIGCOMM]: 6 Mbps, [IIT-B]: 1 Mbps, and [IIIT-D]: 99.5% at 24 Mbps and rest at 1 Mbps. The data rate of PRes's depends on the rate configured in the BSS and type of WiFi clients in the network. Mostly, in 5 GHz lowest data rate is 6 Mbps. Therefore, we find the data rate of PReqs and PRes's in [SIGCOMM] to be 6 Mbps. The presence of legacy WiFi clients mandate the BSS to send PRes's at lowest data rate of 1 Mbps, and for non-legacy WiFi clients, BSS can be configured to send PRes's at a higher data rate. [IIIT-D] has both non-enterprise APs as well as legacy WiFi clients. Thus, 0.5% of PRes's are sent at 1 Mbps.

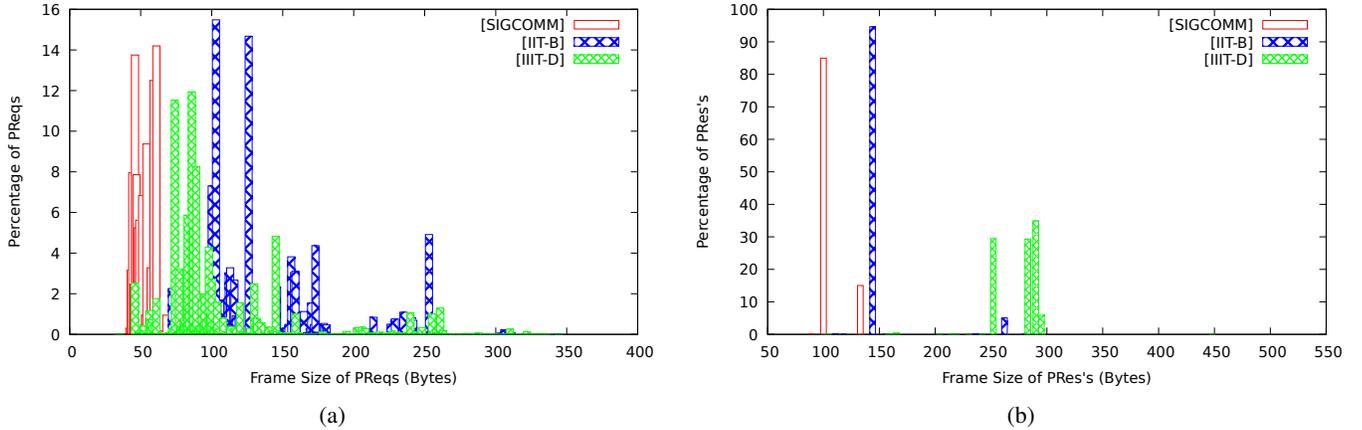

Fig. 7: Size of PReqs and PRes's. We report the percentage of total PReqs or PRes's for each unique frame size. The size of PReqs vary for different WiFi clients, which is not the case for PRes's. Further, [SIGCOMM] has smaller PReqs than the other 2 datasets, while [IIIT-D] has the highest frame size for PRes's.

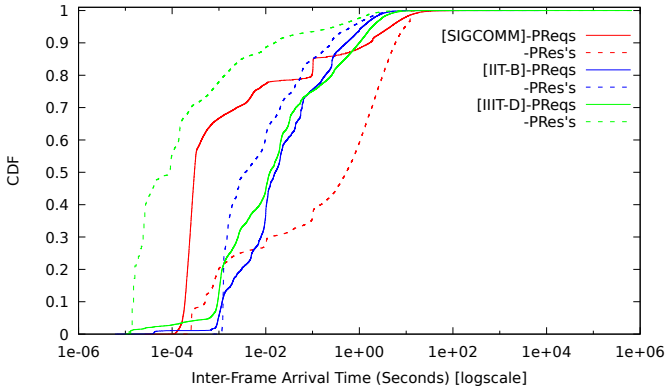

Fig. 8: Inter-Frame Arrival Time (IFAT) of PReqs and PRes's. [IIT-B] and [IIIT-D] see similar IFAT for PReqs. [IIIT-D] sees lowest IFAT of PRes's, followed by [IIT-B].

**Inter-Frame Arrival Times** We define Inter-Frame Arrival Time (IFAT) as the time between 2 consecutive MAC layer frames. Figure 8 shows CDF of IFATs for PReqs and PRes's. For [SIGCOMM], the median IFAT for PReqs is 0.01 ms, whereas for the other 2 datasets it is 10 ms. These numbers show that the WiFi clients were more aggressive in [SIGCOMM] than those in [IIT-B] and [IIIT-D]. This observation conforms with our analysis in Section III. Recall that the average PReqs recorded/minute are 70, 761, and 973 for [SIGCOMM], [IIT-B], and [IIIT-D], respectively. Therefore, even though IFAT of PReqs in [SIGCOMM] is the lowest, least #PReqs/minute will not affect the performance of WLAN, as the clients trigger active scanning in 5 GHz only while they are mobile. Further, the median IFAT of PRes's for [IIIT-D] is less than 0.01 ms, [IIT-B] is less than 10 ms, and [SIGCOMM] is less than 1 sec. A higher #operational BSSs result in reduced IFAT, which is the case of [IIIT-D].

**Airtime Utilization** Airtime Utilization (ATU) is the percentage of time for which the channel is busy. We calculate ATU as proposed in [16]. However, for this study, we quantify this metric separately for data and probe traffic. This analysis allows us to understand, how the growth in probe traffic affects data traffic. Figure 9 shows ATU for [IIT-B]. We do not present ATU of the other two datasets since we observe similar values. Figure 9 shows effect of growth in ATU of probe traffic on the ATU of data traffic. We confirmed this cause and effect in our previous work [2]. Further, we observe that [SIGCOMM] sees the lowest ATU amongst all 3 datasets. On an average, it is less than 0.1% throughout the day. Recall from the analysis presented till now; this dataset has the lowest average #PReqs and PRes's/minute with smallest frame sizes and higher data rates. Furthermore, the #BSS's in [SIGCOMM] is the lowest as compared to other two datasets.

*2) Redundant probe traffic:* Now, we introduce the metric *Redundant probe traffic (RPT)*. We use RPT to compare details of discovered BSS's/scan. Specifically, these details include SSID, BSSID, Channel, and #Associated Clients announced in PRes's for consecutive episodes of active scans. We define an episode of active scan as a group of PReqs separated by less than a second. Given an episode, RPT contains all those PRes's in the current episode that fetch same information about APs as was fetched in the predecessor episode. Figure 10 summarizes the details of RPT for our datasets. We find an average of 90% of PRes's fetch redundant information of nearby BSSs. Precisely, $12926/13069$, $72161/72321$, and $426477/429426$ PRes's are found to carry redundant information in [SIG-COMM], [IIT-B], and [IIIT-D], respectively.

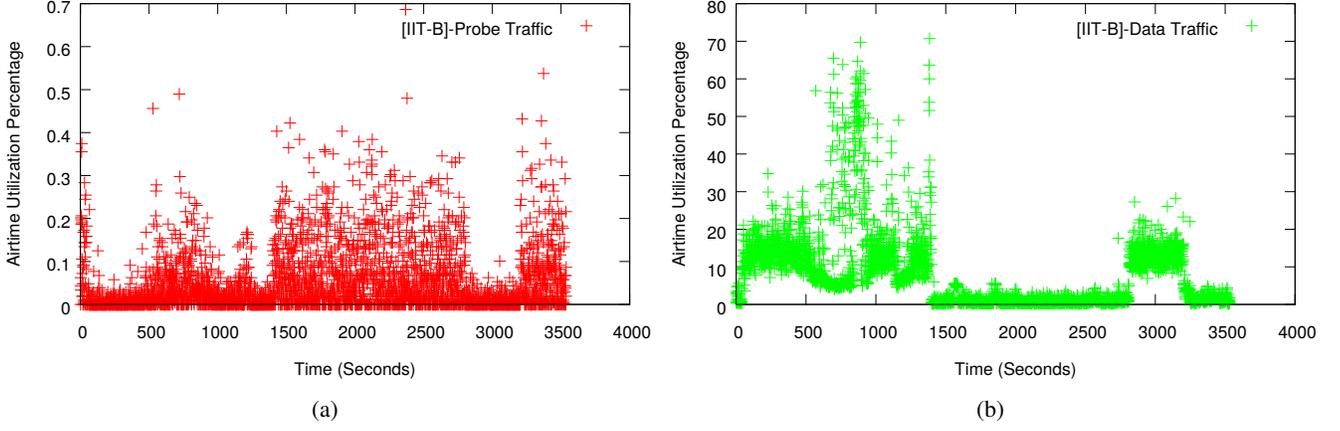

Fig. 9: Airtime utilization (ATU) of probe and data traffic for [IIT-B]. Notice the growth in ATU of probe frames from 1500 to 2700 seconds and the corresponding drop in ATU of data frames in the same period.

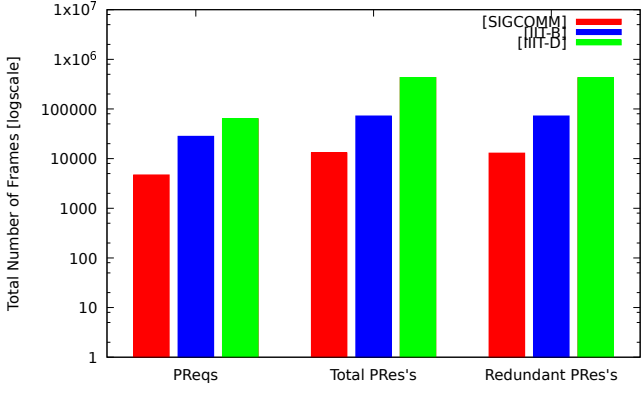

Fig. 10: Total #PReqs, PRes's, Redundant PRes's for [SIG-COMM], [IIT-B], and [IIIT-D]. For all 3 datasets, most of the PRes's carry same information about nearby APs and are hence, shown as redundant PRes's.

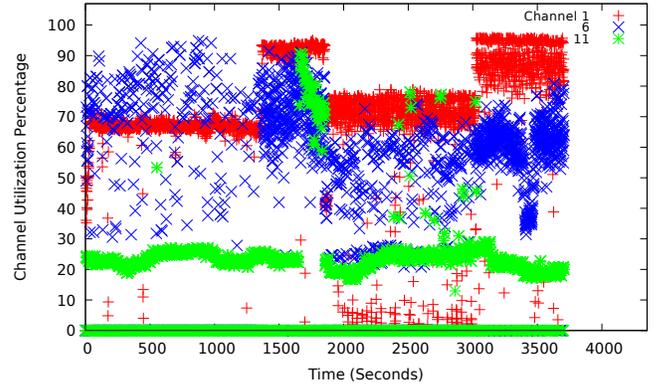

Fig. 11: Channel Utilization as reported by the APs for 2.4 GHz. Notice that channels 1 and 6 are heavily utilized.

We summarize our findings as follows. Not only does active scanning increases latency between a client and its AP, but it does not even help a near-stationary client to find a different AP. On the network-wide scale, frequent active scans inject redundant probe traffic that is low-rate, wastes airtime, and hinders the transmission of data traffic. In a heavily utilized network, these factors negatively impact goodput as we demonstrate in the next section.

### C. Case Study: Effect of Probe Traffic on Goodput

We demonstrate, with a case study, how probe traffic grows in a highly utilized network and ultimately how does that effect the MAC layer goodput. For this purpose, we consider WiFi traffic for an hour on one of the busy days in IIIT-D campus. We collected the traffic with a sniffer configured to listen to 2.4 GHz channels in a round-robin fashion.

Channels were heavily utilized on this day, as shown in Figure 11. We report the Channel Utilization (CU) value defined by `802.11` standard which is measured by the QoS BSS configured in the APs. While a CU higher than 50% denotes heavy utilization of the channel, a CU of 0 signifies either there is no data to be transmitted or clients are not able to transmit due to increased interference and channel contention. We rule out the possibility of no data to be transmitted because we were aware that active WiFi clients were present at the time of data collection. As shown in Figure 11, channels 1 and 6 are continuously heavily utilized, while channel 11 is

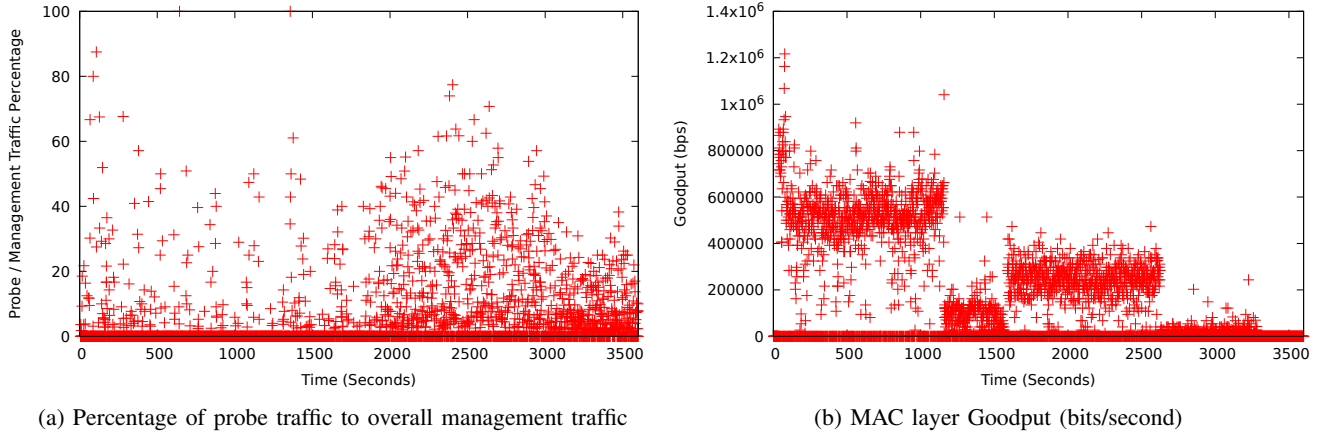

(a) Percentage of probe traffic to overall management traffic

(b) MAC layer Goodput (bits/second)

Fig. 12: Notice the growth in probe traffic at around $1600$ seconds and the corresponding drop[$\approx 50\%$] in the goodput.

intermittently heavily utilized.

Next, we proceed to understand the amount of probe traffic in the network. We measure the percentage of probe traffic with respect to total management traffic recorded every second at the sniffer. Figure 12a shows the results of this measurement. We notice that the probe traffic is sparse before 1600 seconds and after that, it grows drastically. Effect of this growth is visible on the MAC layer goodput, which we demonstrate in Figure 12b. Not only does the goodput drops by more than 50% but even CU increases once the probe traffic grows.

Now that we understand that active scanning severely impacts the performance of WLAN, we discuss ways to reduce probe traffic in the next section.

## V. CONTROLLING PROBE TRAFFIC

In this section, we suggest measures to control the probe traffic from three different perspectives – AP configuration, network planning, and client modification.

Previous section enabled us to understand that most of the APs that could hear a PReq, including the ones sent on overlapping channels, do respond with PRes's. Number of PRes's generated further proliferate with the number of BSSIDs configured per AP. These factors result in the generation of multiple PRes's for a single PReq sent. Moreover, the information transmitted in these responses is redundant, *i.e.*, it does not frequently change as discussed in Section IV-B2.

We propose that the number of operational BSSIDs should be kept as less as possible. Secondly, despite suggested by the WLAN vendors, not all production networks follow the guidelines to reduce PRes's. For example, they don't configure a threshold for the number of PReqs that an AP should respond to, they don't disable PRes's to rogue and unauthorized clients. We presented the observations of these instances in Section IV-A2. In the absence of legacy clients, disable low rate responses. Finally, whenever feasible, enforce 5 GHz operation for the WLAN that can be enabled with band steering.

Above mentioned suggestions are not reliable solutions to control the probe traffic; primarily because they are capable of reducing PRes's in a production network but not PReqs. Furthermore, reducing the number of operational BSSIDs and enabling 5 GHz operation may pose administrative challenges. Therefore, to reduce the PReqs, we suggest to ($a$) improve the planning of a WLAN with respect to the causes that trigger clients to perform active scanning (Section VI) and ($b$) control the amount of PReqs generated at the client-end (Section VII).

## VI. THE CAUSES OF ACTIVE SCANNING AND THEIR INFERENCE

In this section, we answer the question – *"Why do clients trigger active scanning?"* We discuss common causes of active scanning across various devices. We present an inference mechanism to detect them and study their extent in our datasets.

### A. Causes of Active Scanning

The causes of active scanning are broadly grouped into the three categories of *Discovery*, *Connection Establishment*, and *Connection Maintenance*, as shown in Figure 13. We arrive at these causes by empirically analyzing the active scanning behavior of various WiFi clients. Table II summarizes the specifications of these clients. We discuss each of these categories as follows.

*Discovery* At any given time, a client looks for APs in the vicinity at regular intervals. A periodic active scan enables this discovery. Periodic active scans are triggered irrespective of a client's association status and scanning algorithms decide the periodicity. For example, one of the modules in `wpa_supplicant` causes interval to grow exponentially, from a minimum value of 3 seconds to a maximum value of 300 seconds, *i.e.*, 3,9,27,...,300.

*Connection Establishment* A client, if unassociated, looks for suitable APs with which it can establish a new connection and if associated, looks for better APs to which it can roam or handover for a better connection. In either case, a connection is established with a new AP, and this process is a handshake of several messages responsible for association and authentication between the client and the AP. While the process of connection establishment is going on, the client sends PReqs interleaved with association and authentication messages. We believe this is to ensure a faster response from the AP.

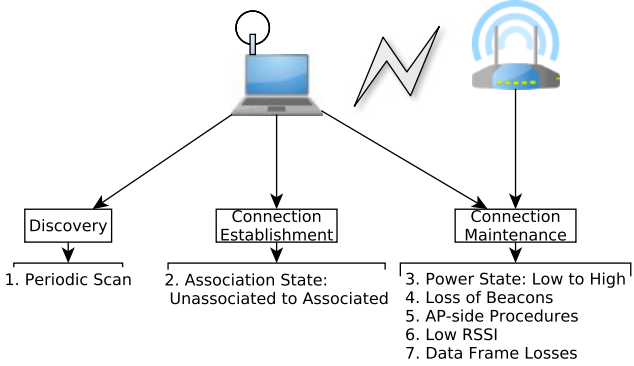

Fig. 13: Causes of Active Scanning

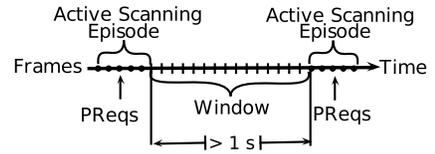

Fig. 14: Splitting a sniffer-based log into active scanning episodes and windows.

*Connection Maintenance* Once connection establishes with AP, its maintenance is done by regularly monitoring the received signal strength from the AP, the fraction of unsuccessfully transmitted data frames, and the rate of beacon loss from the AP. We find active scanning is often coupled with the occurrence of these events. These events correspond to the causes numbered $4, 6, 7$ in Figure 13. Going forward, the procedures at the AP that result in an AP sending deauthentication messages to its clients, trigger active scanning at the clients. Load balancing and band steering are examples of such procedures, which we represent as cause $5$ in Figure 13. Lastly, we have observed active scanning from clients, when they transition to a high power state from a low power state, for example, due to increased user activity. Cause $3$ in Figure 13 represents this cause.

### B. Inference Mechanism for Cause Detection

We propose a sniffer-based inference mechanism for detecting the causes of active scanning. Rationale behind choosing a sniffer is that it is not only agnostic to clients and APs but is easily deployable in any given network.

**Setup** We place the sniffer close to the AP to be monitored. Sniffer is configured to record frames on the same channel as the AP passively. With this setup, the sniffer is likely to record most frames in the network. However, our sniffer-based approach suffers from two limitations. First, sniffer ends up recording frames on overlapping channels [17], which means PReqs sent on overlapping channels will be recorded by the sniffer. Second, overflow at the kernel queues or frames with low RSSI may result in missed frames in the sniffer log. To deal with these limitations, we introduce the notion of *Active scanning episodes* which we explain as follows.

**Approach** We extract frames for each client recorded by the sniffer-based log. We partition these frames into non-overlapping active scanning episodes. Recall that, an episode of active scan is a group of PReqs separated by less than a second. Next, we consider a *window* of frames between two consecutive active scanning episodes, to infer the cause of the episode succeeded by this window. Figure 14 shows an example of an episode and a window. Next, we explain few rules and metrics for inferring the causes.

**Inference Metrics and Rules** The baseline for the proposed metrics and inference rules is knowing the association status of the client. Recall that if the WiFi client is unassociated, we eliminate all the causes under the category of *Connection Maintenance*. We infer the association status of the client with the frames in the sniffer-based log. The absence of frames like data or control indicates that it is unassociated. Otherwise, if such frames are present in addition to the frames that either indicates a successful association, we infer that the client is associated. Further, the presence of deauthentication or disassociation messages from either the client or the AP results in changing the state from associated to unassociated. When we notice frames that cause a client to change its state from unassociated to associated, we infer the cause as *Connection Establishment*. In this case, at least one of association, authentication, or re-association frames precede the active scanning episode. MAC layer header of the captured frames allows us to determine the type of a frame.

As stated earlier, *Connection Maintenance* requires a client to monitor the signal strength from AP, frame drop rates, and beacon loss rates. We infer each of these metrics from the capture in the following ways. Prism/radiotap headers [18], [19] of frames carry the values for signal strength. Since we place sniffer close to AP, the values obtained for signal strength are a good approximation of values seen by the AP as well. We calculate the average and standard deviation for signal strength values and declare the cause as *Low RSSI* if average signal strength is less than -72 dBm and the standard deviation is greater than 12 dB.

To infer the frame drop rates, we measure the number of retried frames, the number of unacknowledged data frames, and reduction in PHY layer data rates. While MAC layer header allows us to know if a frame is a retry, PHY header reveals the data rate of the frames; these values are trustworthy as the client driver sets these values. We estimate the number of unacknowledged data frames by accounting the number of data frames, which are not followed by ACKs. We identify the cause as *Data frame losses* when either or all of the following conditions match – the number of frame retries or ACK losses are more than 50% or PHY layer data rates reduce by more than 50%.

TABLE II: Devices studied to find the causes of active scanning

| Device Type | Chipset Vendor | Operating System | Device Driver |
|---|---|---|---|
| Laptops | Atheros, Intel, and Broadcom | Ubuntu 14.04/12.10 and Windows 8.1/7 | $ath9k$, $iwlwifi$ |
| Tablets | Qualcomm and Broadcom | Android KitKat and iOS 8 | OS provided WiFi drivers |
| Smartphones | Broadcom, Mediatek, and Qualcomm | Android KitKat/Marshmellow/-Jellybean/Cyanogen and Windows | OS provided WiFi drivers |
| USB Adapter | Realtek and Atheros | Ubuntu 14.04/12.10 and Windows 7/8.1 | $rtl8812AU$, $ath9k\_htc$ |

Next cause under this category is *Loss of Beacons*, which we identify by tracking the "awake" messages from a client and inter-frame arrival time of beacon frames. Consecutive null data frames with power save bit set to zero are examples of "awake" messages. In conjunction with these messages, we declare the cause as *Loss of Beacons*, if the inter-frame arrival time for seven beacons in the given window is more than the announced beacon interval, which is 103 ms typically.

Going forward, we identify the cause as *AP-side procedures*, if we encounter deauthenticate message to the client from its AP in the window before an active scanning episode. The last cause under the category of *Connection Maintenance* is *Power State: Low to High*. Note that, we infer the transition of power state for a client and not its power state. The cause is inferred to be *Power State: Low to High* when the number of frames from a WiFi client increases from less than two frames per second to a higher value.

At its end, we infer *Periodic Scan* as the cause, in the absence of an episode mapping to any of the causes discussed above. Note that, this is the only cause that can occur irrespective of a client's association status. All the thresholds mentioned above were derived empirically via experimentation.

Implementation details of the proposed inference mechanism are described in Appendix. It's code implementation, written by us, is open-source and can be found here [20].

### C. Extent of Active Scanning Causes in Real WiFi Deployments

The efficacy of the above-discussed approach is discussed in detail in our previous paper [21]. We used the proposed inference mechanism to analyze the extent of active scanning causes in our three datasets – [SIGCOMM], [IIT-B], and IIIT-D]. We calculate the average percentage for every detected cause over total episodes of active scanning seen in the dataset. Table III shows the result of this analysis.

We summarize the key findings as follows. The *Periodic Scans* by unassociated devices is high for [IIT-B] and [IIIT-D], which means that there are many rogue devices in these networks. The [SIGCOMM] has about 32% due to the cause *Power State: Low to High*, which signifies that user devices were repeatedly changing state from low power to high power and vice-versa. In [IIT-B], about 18% is due to *Low RSSI*. We believe that some clients were distant enough from AP, such that the signal strength was below the threshold. For the other 2 datasets, there was a dense deployment of APs. Data frame losses contribute about 11% in the [SIGCOMM]. We suspect the high number of frame losses is because of a great many clients at the conference venue. For *Connection Establishment*, it is likely that not all devices trigger active scans when attempting to establish a connection. [IIIT-D] has a few number of instances reported for this cause, which makes us guess that probably newer vendor algorithms trigger active scans on connection establishment.

TABLE III: Percentage of occurrence of the different causes of active scanning in three real-world WiFi networks

| Cause | [SIGCOMM] | [IIT-B] | [IIIT-D] |
|---|---|---|---|
| Periodic Scan (Unassociated) | 04.07 | 23.42 | 24.92 |
| Periodic Scan (Associated) | 53.33 | 47.65 | 65.00 |
| Connection Establishment | 00.00 | 00.00 | 00.01 |
| Power State: Low to High | 31.97 | 10.23 | 09.16 |
| Loss of Beacons | 00.00 | 00.00 | 00.03 |
| AP-side Procedures | 00.01 | 00.47 | 00.50 |
| Low RSSI | 00.00 | 18.08 | 00.00 |
| Data Frame Losses | 10.62 | 00.15 | 00.38 |

### D. Reducing Active Scanning with Inference Mechanism

Inferred causes help a network administrator in better network planning. To exemplify, a large percentage of active scanning due to the causes *Low RSSI, Data Frame Losses,* and *Beacon Losses* indicate problems with coverage or excessive interference. These problems get rectified by introducing more APs and ensuring better channel allocation. These measures will mitigate interference due to a high density of APs and clients. *AP-side Procedures* reflect triggering of algorithms at the controller that aim to prevent degradation of network performance. *Periodic Scans* indicate the presence of rogue clients in the network if unassociated, and aggressive clients if associated. The inferred causes enable a network administrator to carefully architect the WLAN deployment. While a well-planned WLAN will indirectly curb the growth of unnecessary active scans, we still need a method to reduce *Periodic Scans*. In the next section, we propose a mechanism to reduce unnecessary active scans from the client-end.

## VII. REDUCING UNNECESSARY ACTIVE SCANS

In this section, we introduce and evaluate the modified scanning strategy to disable unnecessary active scans.

### A. Modified Scanning Strategy

We propose a modified scanning strategy that interleaves active and passive scans. Other than active scanning, clients can passively listen to beacons broadcast by APs to discover nearby APs. This scanning is commonly known as passive scan. Owing to the higher delay than active scanning, passive scanning is a less preferred approach of scanning [22]. However, this decision is taken irrespective of mobility status of a device. The hypothesis behind our proposed strategy is that active scanning is required only when the client is mobile and in all other cases passive scanning should suffice.

As per our proposal, of the 3 principal causes that trigger active scanning, only one cause - *Connection Establishment* should trigger active scans; this includes the case of handover as well.

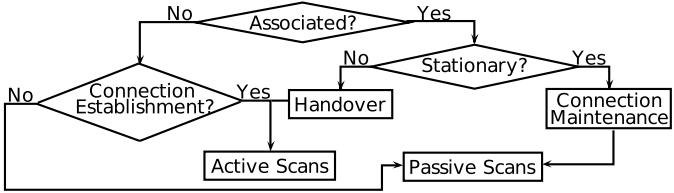

Fig. 15: Modified scanning strategy to decide dynamically decide passive or active scanning on the runtime

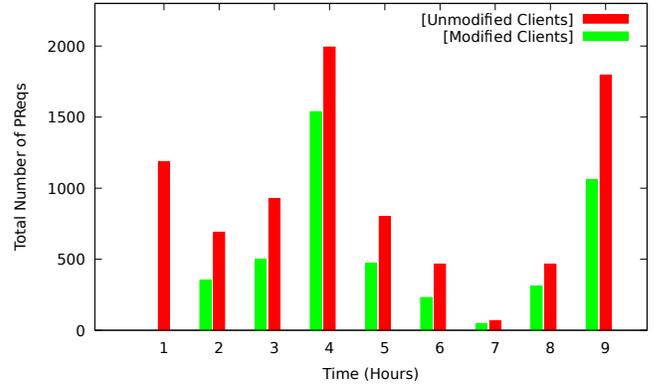

Fig. 16: #PReqs/hour sent by the clients. Notice that in the initial hour, modified clients do not send any PReqs, while there are more than 1000 PReqs from unmodified clients. Even for the subsequent hours, unmodified clients continue to transmit much higher PReqs than modified clients.

Further, as we showed in Section IV 90% of PRes's fetch redundant information about APs, thus active scans due to *Discovery* can be safely disabled for associated clients. For unassociated clients, we propose passive rather than active scans for *Discovery*, reason being unassociated clients do not have any connection-specific strict time constraints

Lastly, during *Connection Maintenance*, a WiFi client needs to make sure that its AP is available. Periodic broadcast of beacons allows easy tracking of AP availability. Thus, passive scanning is sufficient for *Connection Maintenance*.

Figure 15 presents the implementation flow of the proposed approach. A client decides dynamically on the run-time which scan should be triggered. Passive scans are triggered either periodically when the WiFi client is unassociated or when the causes under *Connection Maintenance* occur. Active scans are triggered either during *Connection Establishment* or while handover is happening.

**Effect on Latency** We mentioned earlier that passive scanning is slower than active. Therefore, an argument can be that the proposed scanning scheme might add significant latency to an ongoing connection. However, this will not be the case because firstly periodic scanning which is the most common cause (Average 72%), either passive or active, is altogether disabled and secondly, this scheme triggers passive scans only during *Connection Maintenance* while the client is immobile. Thus, passive scanning is triggered only in necessary conditions like *AP-Side Procedures* and *Data Frame Losses*. Unless such conditions are triggered, which is unlikely in a well established WLAN, passive scanning will not trigger; thereby preventing unnecessary latency at the client.

### B. Evaluating the Modified Scanning Strategy

*1) Code Implementation:* We instrumented Ubuntu 16.04 machines to implement the above-discussed scanning strategy. All kinds of network managers were disabled on the machines. We modified `wpa_supplicant` [23] code to decide at runtime whether to trigger active or passive scans. This was enabled with the `bg_scan`'s `simple` module [24]. By default, we enabled passive scans, and disabled active scans. Active scanning was enabled only when the RSSI value as seen by the client from AP dropped below −70 dBm, which is an empirically derived estimate. We did not control the PReqs generated by `mac80211` code or lower level device driver code since that will defeat our purpose of developing a vendor-neutral solution.

*2) Experiment Setup:* We performed a 9-hour long experiment to evaluate the performance of the modified scanning strategy in a real WLAN of IIIT-D. In addition to enterprise-grade APs of IIIT-D WLAN, we installed two more APs that we could control. Both the APs had an overlapping coverage, broadcasted the same SSID, and were on the same channel. This configuration was to enable smooth handover for clients. We collected the sniffer logs with TP-Link WN721N sniffer placed close to the AP to which clients were associated. Sniffer passively listened on the same channel as of the AP. We had 20 WiFi clients, 10 with modified scanning code and 10 with unmodified scanning code. They had WiFi chipsets from Atheros, Broadcom, MediaTek, Ralink, and Realtek.

We begin with all the devices in the unassociated state. After an hour, we associated them to one of the APs. Every hour, we randomly introduced the causes that trigger active scanning. Each client was instrumented to log the timestamps of various connection related events such as association, authentication, scanning started, scanning ended, etc. Note that this setup was created to simulate a real WLAN. We did not control any other network or client parameters.

**Hypothesis** We hypothesize that modified clients will induce lesser PReqs than unmodified clients without affecting their connection.

*3) Results:* We evaluate the performance of the clients for #PReqs transmitted, the duration for which their connection persisted [3], and the time it takes to re-establish a broken connection.

**#PReqs** Figure 16 shows #PReqs as recorded by the sniffer. During the first hour, when all the clients were unassociated, unmodified clients transmitted 1184 PReqs while there were no PReqs from modified clients. Since WLAN setup does not change throughout the experiment, all these frames induced RPT in the network. For the following hours, unmodified clients always transmitted higher PReqs than modified clients. Modified clients sent a maximum of 50% and a minimum of 22.61% lesser PReqs than the unmodified clients.

**Network Discovery** Clients under observation were near-stationary throughout the experiment except during the induc-tion of the cause *Low RSSI*. Therefore, the APs discovered at the clients remain to be the same. Our modified scanning strategy deliberately enables active scanning on *Connection Establishment*, so there is a possibility of inherent switching delay that occurs due to the transition from passive to active scanning. Thus, we analyze the delay to connection setup by analyzing the time elapsed between disconnection and connection. We refer to this as Time To Connect (TTC).

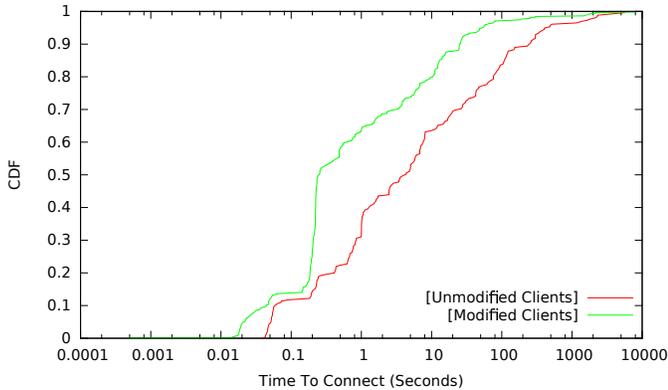

Fig. 17: Time to connect. Median time to connect for unmodified clients is $4.35$ seconds, whereas for modified clients it is $0.25$ seconds. Even $75^{th}$ percentile for unmodified clients is $43$ seconds, whereas for modified clients it is $5.7$ seconds.

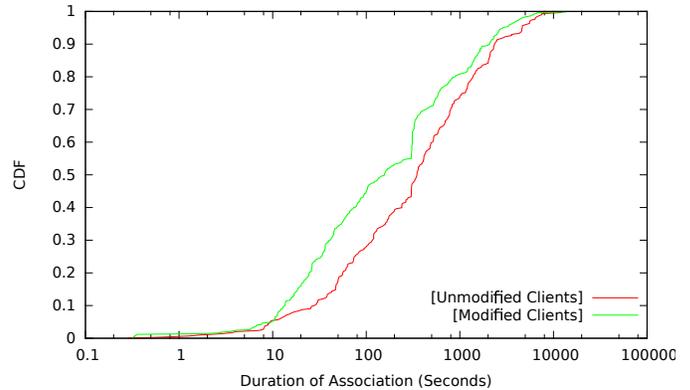

Fig. 18: Connection persistence. Notice that modified clients continue to stay connected with higher probability. Modified clients show the maximum duration of association being $1196$ seconds higher than unmodified clients.

**Time To Connect** We calculate TTC by analyzing the timestamps recorded for various events at the clients. Typically, when a disconnection is triggered series of events as recorded at the client-end are disconnected, scanning, and connected. In our case, the scanning can be either active or passive. The difference of timestamps recorded for disconnected and connected events give us the value of TTC. Figure 17 presents the CDF of TTC. We observe that for the modified clients TTC is much lesser than the unmodified clients. In fact, at the $50^{th}$ percentile the TTC is mostly constant at approximately $0.25$ seconds for modified clients, whereas for unmodified clients it goes up to $4.35$ seconds. The reason for much lower TTC is that the modified clients are not involved in unnecessary active scans and hence, its transmission queues are relatively less loaded.

**Connection Persistence** Once the client is connected, transmitting lesser PReqs does not deter the performance of WiFi connection. We confirm this with our next metric that measured the duration of connection persistence. We measure this duration with the time elapsed between 2 successful connection establishments as recorded at the client. For the entire length of experiment, modified clients persistently showed better connection persistence, as shown in Figure 18. We investigate the probabilities for $1$, $5$, and $15$ minutes of association. While the unmodified clients demonstrate probabilities as – $0.20$, $0.43$, and $0.73$, the modified clients remained associated with probabilities – $0.37$, $0.55$, and $0.80$. Clearly, the modified clients demonstrate higher chances of staying associated than the unmodified clients.

In fact, when we induced causes such as *Data Frame Losses* and *AP-side Procedures*, driver code of few unmodified froze the complete system, and we had no other option than to restart the machine. This behavior shows that unmodified clients had the intolerance towards few causes. This behavior is one of the reasons behind lower association duration of unmodified clients.

With our scanning strategy, we successfully demonstrated that the stationary clients need not perform unnecessary active scans. Even without it, the WiFi connection did not suffer, but it was maintained in a much better manner. Last but not the least, implementation of the proposed strategy can be easily rolled out to devices as a simple application update. As a limitation of this strategy, stationary clients will experience higher delays to ongoing connection if losses increase at alarming rates.

## VIII. RELATED WORK

In this section, we summarize existing works that study the procedure, the extent and the impact of active scanning in WLANs. We briefly, outline the solutions developed to solve the problems due to active scanning. Towards the end we talk about the metrics and the systems available to improve the performance of WLANs.

**Procedure and Extent of Active Scanning** The procedure of active scanning is well-studied in literature. Authors in [25]–[27] empirically explain the sequence of events that happen during active scanning. These papers present the intrinsics of active scanning from the perspective of a client. Authors in [25] empirically study the exchange of PReqs and PRes's and analyze their impact on handover process. Hu et al. in [28] present the dynamics of PReqs in large-scale and dense WLANs. Their study covers venues like stadiums and classrooms. They quantify the details of PReqs across frequency bands, scan interval, scan duration, and the number of SSIDs.

**Impact of Active Scanning** The impact of active scanning on the performance of a client concerning increased delay during handover process is studied in detail [5]–[9], [11], [25], [26]. Its impact on the energy expenditure of a client is examined by

Hu et al. in [28] and Raghavendra et al. in [3]. They recognize the negative impact of PReqs on the performance of a dense WLAN. They describe the issues with current procedures of active scanning and how they are detrimental to the operation of a WLAN. The analysis presented in these studies is either limited to client-side, or it partially covers the aspects of probe traffic that affect performance on a network-wide scale. In Section IV, we extended this analysis by covering various aspects of probe traffic – the amount, the inter-frame arrival times, frame sizes, data rates, airtime utilization, and redundant probe traffic.

**Solving the Problems from Active Scanning** The state-of-the-art has solutions for a WiFi client to discover APs with reduced scanning delays. The solution proposed in [5] suggests synchronized scanning among APs and clients. DeuceScan [6] exploits spatiotemporal information available at a client to find the best AP to roam to. FastScan and Enhanced FastScan [7] suggest the use of a client-side database of APs. Authors in [8] propose an interleaving technique, to use an optimal combination of active/passive scans. Multiple radios [9] along with recently developed Multipath Transmission Control Protocol [10] allows a client to stay associated with multiple WiFi networks via virtual interfaces. CSpy [11] finds the strongest channel using Channel Frequency Response and thereby sending PReq on only one channel. It then predicts the quality of nearby channels Authors in [25] and [26] suggest measures to reduce the time spent in active scanning.

The majority of these solutions aim to reduce the handoff delays rather reducing unnecessary PReqs and PRes's. Some of these solutions are proposals with possible amendments without any implementation details. Others either provide driver-specific or hardware-specific solutions. Such solutions are not device agnostic and thus, can not be implemented across a broad range of WiFi clients. In our scanning strategy, we hypothesize that by reducing the number of PReqs, we are indirectly disabling the generation of unnecessary PRes's from APs. It takes care of the active scanning triggers for stationary clients. It is an amendment in the existing `wpa_supplicant` code.Therefore, it can be simply rolled out as an update to Linux and Android systems without worrying about the underlying chipset.

**Metrics and Systems for Measuring and Improving the Performance of Network** Literature suggest several metrics, to gauge the performance of a network. We use few of these metrics to measure the impact of active scanning on clients in specific and on the complete network in general. First of these metrics is latency. Latency is an important parameter that allows measuring the impact on real-time applications such as voice and games [12]. We used this metric to analyze the client-side impact of active scanning. Next, we used inter-frame arrival time [28] to study how frequently PReqs and PRes's arrive in WLANs. We used airtime utilization [16] to quantify and compare the amount of airtime consumed by probe and data traffic. This metric enabled us to understand that for a given channel, data traffic is bound to suffer if probe traffic consumes significant airtime. Lastly, we used connection persistence [3] to measure the durability of a WiFi association. We used this metric to analyze the performance of our modified scanning strategy.

Detecting the problems in large scale WiFi networks through sniffer-based logs is common in literature, partly because such solutions are device agnostic and therefore vendor neutral. PIE [29] and Jigsaw [30] rely on capturing the WiFi traffic through passive sniffing. These systems are used to detect potential problems in a WLAN, in particular, that of interference. We use these systems as our motivation for the proposed inference mechanism.

## IX. CONCLUSION

We empirically demonstrated that WiFi clients trigger unnecessary active scans. Such scans inject low-rate probe traffic in the network that eventually results in degraded goodput. We showed this phenomenon is more relevant to $2.4$ GHz than $5$ GHz. We suggested three ways to control the growth of probe traffic in production WLANs. We experimentally showed the applicability of two of these ways in real WLANs.

## ACKNOWLEDGEMENT

Authors would like to thank LiveLabs, SMU for kindly agreeing to provide WiFi traffic used to compare the scanning behavior of devices in $2.4$ and $5$ GHz.

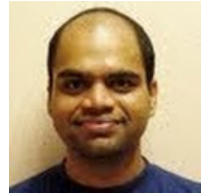

**Vinayak Naik** He is an associate professor at IIIT-D, a Research Advisor to TCS, and a Chief Architect at BackPack. He spent summer of 2016 as a visiting researcher with MNS group at MSR, Bangalore and summer of 2015 at VMware R&D. At MSR, he worked on IoT projects. At VMware, he worked on SDN and NFV. In the past, his work has transitioned into industry. The examples are ExScal project to Northrop Grumman and Samraksh, Saarthi project to DIMTS, and RighFare to Suruk.

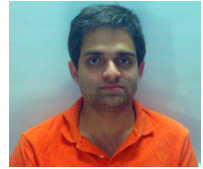

**Sanjit Kaul** He is assistant professor at IIIT-D, since October 2011. He received a PhD. in Electrical and Computer Engineering from Rutgers University in 2011. His research was in the area of improving driver safety amongst large networks of on-road vehicles, which he pursued at Wireless Information Networks Laboratory (WINLAB). Before joining graduate school he worked on 3G systems at Hughes Software Systems, now Aricent, and Ubinetics, now CSR, India, for about four years.

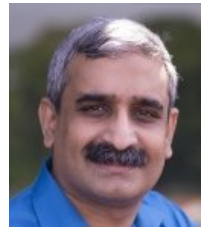

**Rajesh Balan** Rajesh is an associate professor at Singapore Management University's School of Information Systems. He received his Ph.D. in computer science from Carnegie Mellon University and has over 15 years of research experience in the broad area of mobile systems and software. Some of the diverse areas that he has worked on include infrastructure support for multiplayer mobile games, improvements to public transportation networks, understanding and improving the software development process in outsourced environments, power management for mobile displays and sensing, and developing and testing novel human-centric mobile applications. Rajesh is also a director of the LiveLabs Urban LifeStyle Innovation Platform. The goal of this platform is to allow mobile sensing, analytics, applications and services to be tested with real users on real phones in real-world environments. Currently, LiveLabs has been deployed at a university campus, a resort island, and a large convention centre. More details about LiveLabs can be obtained at http://www.livelabs.smu.edu.sg.

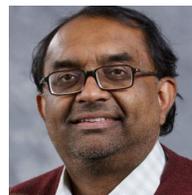

**Sumit Roy** He has been a faculty member in Electrical Engineering at the University of Washington since 1998, where he is presently Integrated Systems Professor. He spent 2001-03 on academic leave at Intel Wireless Technology Lab as a Senior Researcher engaged in systems architecture and standards development for ultra-wideband systems and next generation high-speed wireless LANs. During Jan-Jul 2008, he was the Science Foundation of Ireland's E.T.S. Walton Awardee for a sabbatical at University College, Dublin, and during summer 2011 he was the recipient of a Royal Academy of Engineering (UK) Distinguished Visiting Fellowship. Roy's activities for the IEEE Communications Society (ComSoc) include membership of several technical and conference program committees. He has served as Associate Editor for all the major ComSoc publications in his area at various times, including the IEEE Transactions on Communications and IEEE Transactions on Wireless Communications, and as a Distinguished Lecturer (2014-15) for ComSoc. He was elevated to IEEE Fellow by ComSoc in 2007 for "contributions to multi-user communications theory and cross-layer design of wireless networking standards."

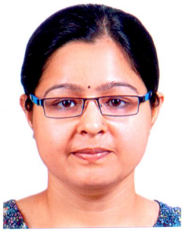

**Dheryta Jaisinghani** She is a PhD research scholar at Wireless Systems Lab, IIIT-Delhi, India. She is advised by Dr. Vinayak Naik and co-advised by Dr. Sanjit Kaul. Her areas of interest are large-scale and dense WiFi networks, such as enterprise scale WiFi networks, mobile, and ubiquitous computing. Prior to joining IIIT-Delhi, she graduated with gold medal in M.Tech from IIIT-Bangalore in 2012. During master's she was advised by Dr. P. G. Poonacha (IIIT-Bangalore). She has an industry experience with Avanade Technology Group, Accenture Services